# Evidence that self-similar microrheology of highly entangled polymeric solutions scales robustly with, and is tunable by, polymer concentration

Ian Seim and Jeremy A. Cribb

*Department of Mathematics* and *Department of Physics and Astronomy, University of North Carolina at Chapel Hill, Chapel Hill, North Carolina 27599*

Jay M. Newby

*Department of Mathematics, University of North Carolina at Chapel Hill, Chapel Hill, North Carolina 27599*

Paula Vasquez

*Department of Mathematics, University of South Carolina, Columbia, South Carolina 29208*

Martin Lysy

*Department of Statistics and Actuarial Science, University of Waterloo, Waterloo, Ontario N2L 3G1, Canada*

M. Gregory Forest and David B. Hill

*The Marsico Lung Institute, Departments of Applied Physical Sciences, Mathematics, Physics and Astronomy,* and *Biomedical Engineering, University of North Carolina at Chapel Hill, Chapel Hill, North Carolina 27599*

# Abstract

We report observations of a remarkable scaling behavior with respect to concentration in the passive microbead rheology of two highly entangled polymeric solutions, polyethylene oxide (PEO) and hyaluronic acid (HA). This behavior was reported previously [Hill *et al.*, *PLOS ONE* **9,** e87681 (2014)] for human lung mucus, a complex biological hydrogel, motivating the current study for synthetic polymeric solutions PEO and HA. The strategy is to identify, and focus within, a wide range of lag times $\tau$ for which passive micron diameter beads exhibit self-similar (fractional, power law) mean-squared-displacement (MSD) statistics. For lung mucus, PEO at three different molecular weights ($M_w$), and HA at one $M_w$, we find ensemble-averaged MSDs of the form $\langle \Delta r^2(\tau) \rangle = 4\, D_\alpha \tau^\alpha$, all within a common band, [1/60 sec, 3 sec], of lag times $\tau$. We employ the MSD power law parameters ($D_\alpha, \alpha$) to *classify* each polymeric solution over a range of highly entangled concentrations. By the generalized Stokes-Einstein relation, power law MSD implies power law elastic $G'(\omega)$ and viscous $G''(\omega)$ moduli for frequencies $1/\tau$, [0.33 sec$^{-1}$, 60 sec$^{-1}$]. A natural question surrounds the polymeric properties that dictate $D_\alpha$ and $\alpha$, e.g., polymer concentration *c*, $M_w$, and stiffness (persistence length). In [Hill *et al.*, *PLOS ONE* **9,** e87681 (2014)], we showed the MSD exponent $\alpha$ varies linearly, while the pre-factor $D_\alpha$ varies exponentially, with concentration, i.e., the semi-log plot, $(\log(D_\alpha), \alpha)(c)$, of the classifier data is collinear. Here we show the same result for three distinct $M_w$ PEO and HA at a single $M_w$. Future studies are required to explore the generality of these results for polymeric solutions, and to understand this scaling behavior with polymer concentration.

**Introduction**

Highly entangled polymeric solutions are ubiquitous in biology and materials science. Their macroscopic and microscopic rheology exhibit the distinct signature of self-similarity in the scaling of viscoelastic moduli over a significant frequency range, or in the creep compliance over a broad range of times, or in the MSD statistics of passive micron-scale beads over a broad range of lag times.  Self-similar rheology is not only intriguing, it allows one to extrapolate viscous and elastic moduli across the frequency and timescale range of power law behavior.  Here we present results for 4 different polymeric solutions (polyethylene oxide at three $M_w$ and hyaluronic acid at one $M_w$) for which the power law microbead rheology not only persists for each highly entangled concentration, but the power law parameters have a remarkable, simple scaling with concentration.   We are not aware of any theoretical basis for this behavior, and hope these results will spawn interest of both experimentalists to explore different polymers and theorists to understand this scaling behavior.

Fractional, or power law, or self-similar rheological properties of entangled polymer solutions have led to the development of new classes of rheological models at the continuum (macrorheological) and microscopic (microrheological) scales, shifting from traditional exponential kernels [1, 2] to generalized Rouse exponential series that yield tunable power laws [3, 4] and explicit fractional or power law kernels [1, 5-12].  These power law models for both microrheology and macrorheology have a small number of parameters, e.g., two or three, instead of orders of magnitude larger parameter sets with traditional exponential kernels, rendering inverse characterization from experimental data tractable.  From such power law models, powerful statistical tools are amenable to experimental data:  Bayesian model selection methods choose among candidate models, cf. [13] for macrorheology and [4] for microrheology; and data-based maximum likelihood estimation (MLE) of parameters for the best-fit model; cf. [14] for macrorheology and [15] for microrheology.

In the present paper, we use microbead particle tracking experiments (cf. [14] and data analytics tools [4, 15], focusing exclusively within a self-similar range of lag times / frequencies, using the power law MSD parameters ($D_\alpha,\alpha$) as a *classifier* for individual, highly entangled polymeric liquids. Our shortest experimental lag time is 1/60 sec, and longest lag time with sufficient observational data is 3 sec. *Within* this range, tracked microbeads in each solution considered here exhibit self-similar, power law MSD behavior, and by inference, self-similar microrheology within a limited yet significant frequency range from the generalized Stokes-Einstein relation, cf. [16, 17]. We acknowledge that this classifier is not a full characterization of polymeric solutions, nor is it intended to be.

For many biopolymeric solutions, notably mucus barriers in the human lung, gut, and female reproductive tract, practical limitations, including low volume availability and very low yield thresholds, have led to passive particle tracking (PPT) microrheology as the most viable, robust, and reproducible assessment of rheological properties [18-20]. Typical PPT experiments record the position time series of ensembles of dispersed 0.2-2$\mu$m diameter beads at 50-100 frames per second for sufficiently long durations; 30 second duration at 60 frames per second is typical for mucus experiments of the authors [4, 14, 15]. The squared increments (displacements), $\Delta r^2(\tau)$, versus lag time, $\tau$, of the particle position time series are typically ensemble-averaged to yield a summary statistic, the mean-squared displacement (MSD=$\langle \Delta r^2(\tau) \rangle$), for each highly entangled polymeric liquid sample.

The MSD summary statistic, when self-similar over a range of lag times, provides a two-parameter basis for comparison of each experimental sample with other soft materials. The Fourier (or Laplace) transform of the MSD, together with the generalized Stokes-Einstein relation [16, 17], yields an immediate expression for the linear dynamic moduli,

$$G^*(\omega) = \frac{k_B T}{\pi a i \omega \Im(\langle \Delta r^2(\tau) \rangle)} \qquad (1)$$

where $a$ is the particle radius, $\Im(\cdot)$ denotes the Fourier transform, and $\langle \Delta r^2(\tau) \rangle$ is the MSD statistic. When the MSD statistic admits a power law scaling $\langle \Delta r^2(\tau) \rangle = 2d\, D_\alpha \tau^\alpha$, where $d$ is dimensionality, relating lag time and frequency, $\tau = \frac{1}{\omega}$,

$$\langle \Delta r^2 \left(\frac{1}{\omega}\right) \rangle = 2dD_\alpha \left(\frac{1}{\omega}\right)^\alpha ; \qquad (2)$$

by the results of Mason in [17], we have

$$G^*(\omega) = \left[\frac{k_B T \cos\left(\frac{\alpha \pi}{2}\right)}{\pi a \Gamma(1+\alpha) D_\alpha}\right] \omega^\alpha + i \left[\frac{k_B T \sin\left(\frac{\alpha \pi}{2}\right)}{\pi a \Gamma(1+\alpha) D_\alpha}\right] \omega^\alpha = G'(\omega) + iG''(\omega), \qquad (3)$$

where $\Gamma(\cdot)$ is the Gamma function. The numerical pre-factor of 2*d* in the MSD is chosen for compatibility with fractional Brownian motion (fBm) as a model for power law MSD scaling. The MSD formula (2) is *exact* for fBm; furthermore, we have high-precision inverse tools for determination of the pre-factor $D_\alpha$ and exponent $\alpha$ from PPT data, even in the presence of deterministic particle drift [4,17] that is evident in previously published HBE mucus data [16] as well as the PEO and HA data analyzed here. The advantages of coupling a model for sub-diffusive motion in the presence of drift are discussed in detail in [17]. The upshot is that power law scaling of MSD in the time domain is equivalent to power law scaling of $G'$ and $G''$ in the frequency domain. Furthermore, it immediately follows that the polymeric solution properties where $\alpha = 1/2$, if accessible within the self-similar range, predict a transition where the viscous and elastic moduli "cross over" from viscous- to elastic-dominated [21]. We note that a similar expression can be found using Fourier transform and fractional calculus as discussed in [22].

In human bronchial epithelial (HBE) airway mucus, a proxy for polymeric concentration is the weight percent (wt%) of solids since HBE mucus contains extremely high $M_w$ mucin species, together with diverse protein crosslinks. In recent PPT experiments spanning wt% solids from health to disease [23], a remarkable scaling relation is revealed in the ensemble-averaged MSD statistic versus HBE mucus concentration. First, we found for each fixed concentration of HBE mucus, PPT data (after drift has been removed by methods in [4,17]) exhibits a power law MSD over the entire range [1/60 s, 3 s] of lag times $\tau$, $\text{MSD}(\tau) = 2d\, D_\alpha \tau^\alpha$. (We note that after using maximum likelihood estimation of drift and fBm parameters to individual microbead position time series, by direct numerical simulation of fBm + drift, we are able to reconstruct each individual microbead trajectory.) The self-similar, fractional behavior is not surprising, since HBE mucus is a highly entangled, cross-linked polymeric solution at physiological concentrations. The experimental data (per tracked microbead and then ensemble-averaged), yield mean and standard deviation of $\alpha$ and $D_\alpha$ for each fixed concentration *c*. When analyzed over a broad concentration range, we find that $\alpha$ is linearly proportional to *c*, while $\mathcal{D}$, a non-dimensional $D_\alpha$ defined in the Appendix, scales exponentially with *c*. (N.B. We non-dimensionalize $D_\alpha$ to $\mathcal{D}$ for an important reason. One cannot compare effective diffusivities $D_\alpha$ except at constant $\alpha$, since the physical units scale with $\alpha$, as is evident from the MSD formula. With $\mathcal{D}$, just like with $\alpha$, one can make relative comparisons of mobility, just as one would compare diffusivities of viscous fluids. Furthermore, this non-dimensionalization does not affect the scaling behavior with respect to concentration that is the central focus of the paper.)

Thus, the experimental data versus c is statistically collinear if plotted on the semi-log scale $(\log(\mathcal{D}), \alpha)$ versus c; see Figure 1. This result can be restated as follows: *c* provides a parametric representation of a robust linear relationship between $\alpha$ and $\log(\mathcal{D})$ over a health-to-disease range of *c* for HBE mucus. The slope and intercept of this linear relation are apparently determined by other HBE mucus properties that are evasive given the spectrum of mucin glycoproteins and other proteins that crosslink mucins, as well as salts in HBE mucus. We note that $\alpha$ and $\log(\mathcal{D})$ decrease linearly with increasing *c*, implying increasingly sub-diffusive motion (reduced mobility with stronger correlations in the increments) of passive microbeads in HBE mucus as wt% solids increases (notoriously associated with increasing disease pathology). These results, when translated by formula (3) to dynamic moduli that are proportional to $1/D_\alpha$, reveal *orders of magnitude shifts* in $G'$ and $G''$ as HBE mucus concentration shifts from normal concentrations to the disease range, shown in Figure 2. Furthermore, there is a transition near 4 wt%, where the elastic and viscous moduli "cross over": HBE mucus transitions from a viscoelastic fluid at low wt% solids to a viscoelastic solid above 4 wt% solids. We further find at the cross-over transition that the ensemble MSD data is significantly more heterogeneous (see [23] for details), yielding the largest departure from the best-fit line for the HBE mucus data and correspondingly larger error bars.

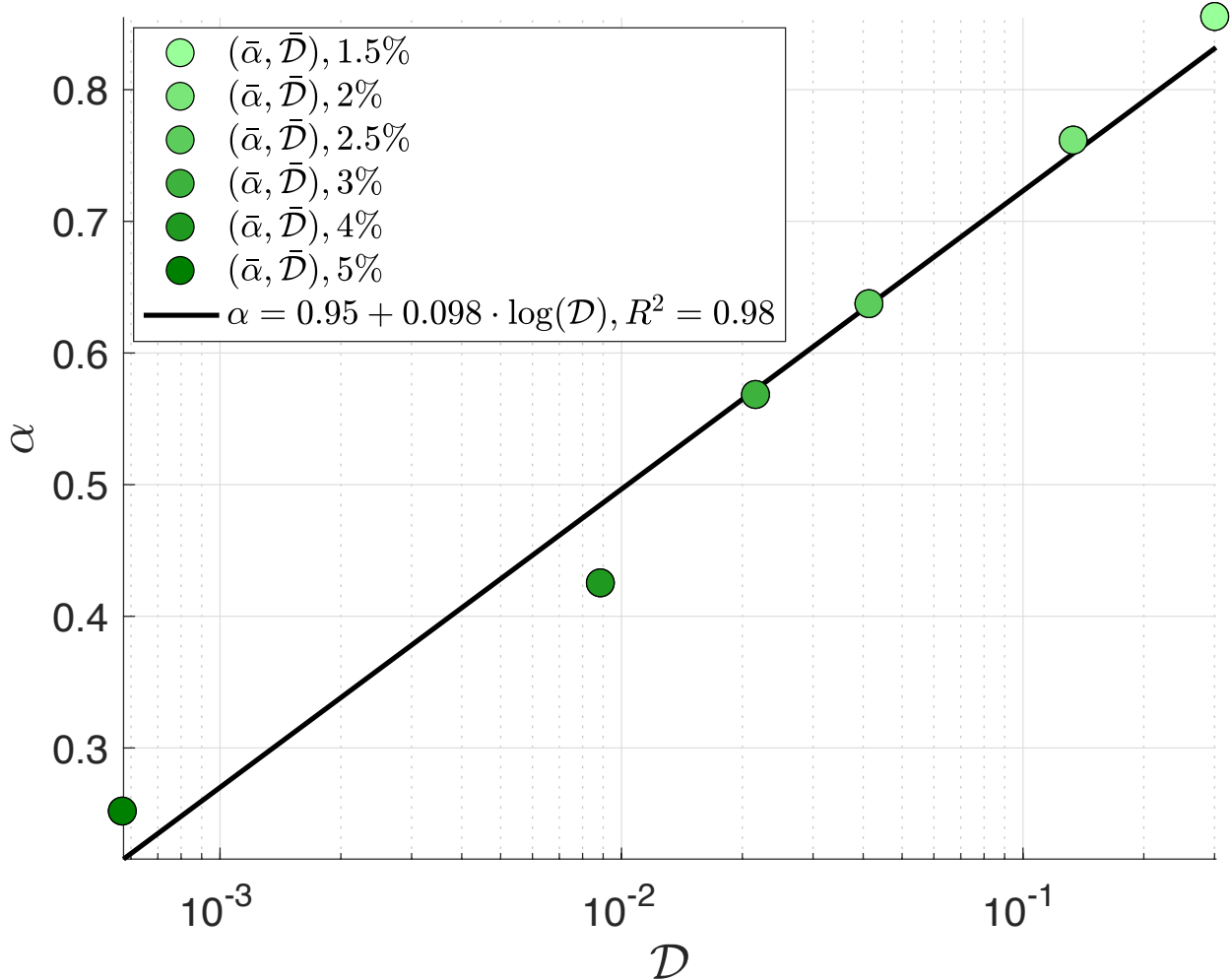

**Figure 1**: Passive particle tracking data [23] from HBE cell culture mucus over a range of wt% solids. Data correspond to the ensemble-averaged means of MSD parameters ($\log(\mathcal{D})$ and $\alpha$) for HBE mucus concentrations spanning healthy (1.5-2.0% solids) to pathological (4-5% solids). The methods for calculating the lognormal distribution of $\mathcal{D}$ and the normal distribution of $\alpha$ are detailed in the Data Analytics section.

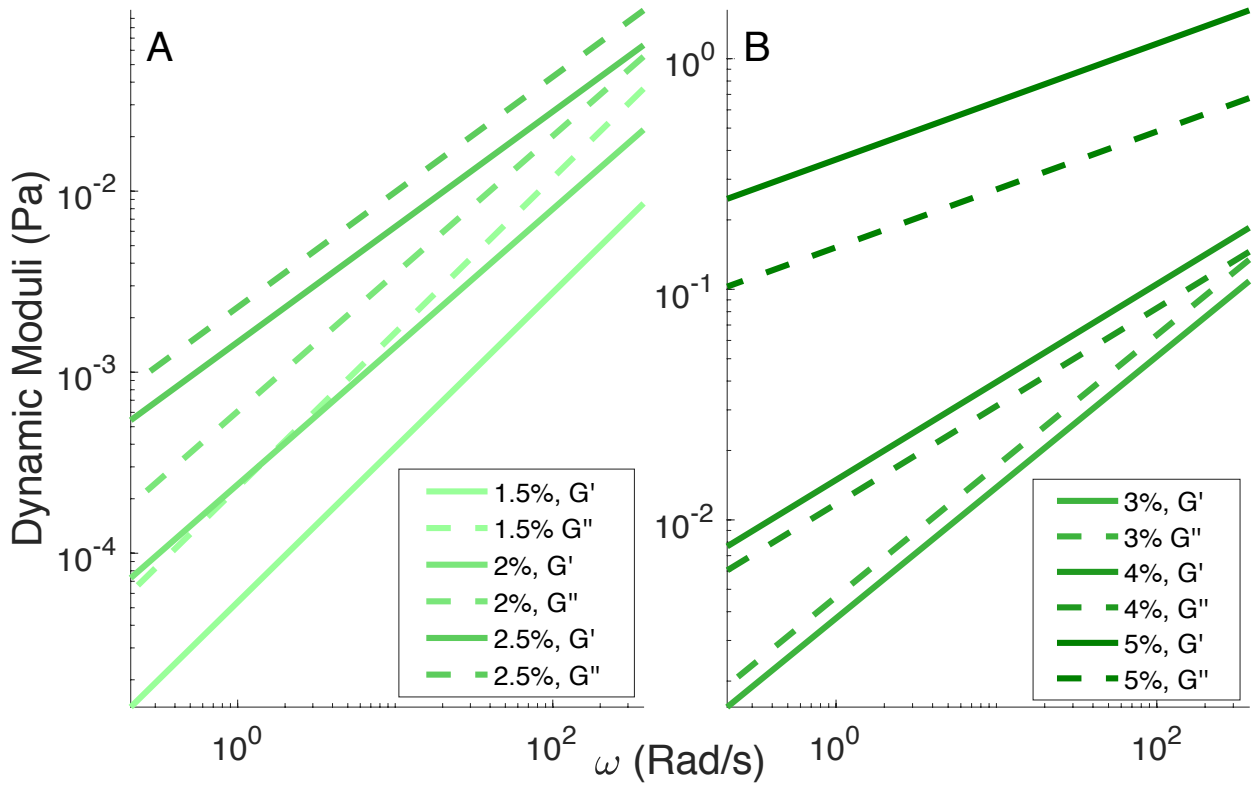

**Figure 2**: Dynamic viscoelastic moduli, $G'$ and $G''$, for HBE mucus samples that follow directly from the ensemble MSD self-similar scaling behavior of Figure 1 [23]. These plots are indistinguishable when applying formulas in (3) versus formula (1) with ensemble, drift-removed, MSD experimental data.

As noted earlier, the HBE mucus scaling results in Figures 1 and 2 lead to a natural *hypothesis*: the self-similar microrheology of highly entangled polymeric solutions scales primarily with polymer mass concentration, while other polymer properties such as molecular weight and persistence length dictate the range of viscous and elastic properties swept as concentration varies. That is, we hypothesize that PPT experimental data, $(\log(\mathcal{D}), \alpha)$ versus $c$, are collinear in the highly entangled regime, while other polymer properties determine the slope and intercept of the polymer-specific line. Since HBE mucus is a mixture of several large molecular weight mucin macromolecules together with potentially scores of crosslinking protein species, DNA, and other cellular debris, we decided to explore this hypothesis with two synthetic polymer solutions. The first is polyethylene oxide (PEO), a synthetic, flexible, non-cross-linked, relatively homogeneous, polymer solution. We employ PEO at three different $M_w$, 1, 5, and 8 Mega Daltons (MDa), and for each $M_w$, over a range of concentrations well above the overlap concentration, in the highly entangled regime. We repeated the same PPT microrheology experiments carried out for HBE mucus [23] with $1\mu m$ diameter beads, using similar data analytics tools described in [15]. In Figure 3, we present the results for 1, 5, and 8 MDa PEO for *c* in the highly entangled regime. Transitions from this linear scaling at *c* below the entanglement concentration are presented in Figure 4.

    Figure 3 shows for each $M_w$ PEO, a remarkable collapse of the ensemble-averaged MSD statistic versus concentration, $\text{MSD}(\tau) = 4\,D_\alpha \tau^\alpha$, where again we replace the dimensional pre-factor $D_\alpha$ by $\mathcal{D}$ (see Appendix) so that both mobility parameters admit relative comparisons. *The hypothesis is quite strongly confirmed for all three PEO solutions: the $(\log(\mathcal{D}), \alpha)$ versus c data are collinear with a coefficient of determination $(R^2)$ of .99.* We further observe that both the slopes and the intercepts drop in an apparent linear fashion on a $\log M_w$ scale (see Appendix). The results for very high $M_w$, multi-species HBE mucus are included for comparison.

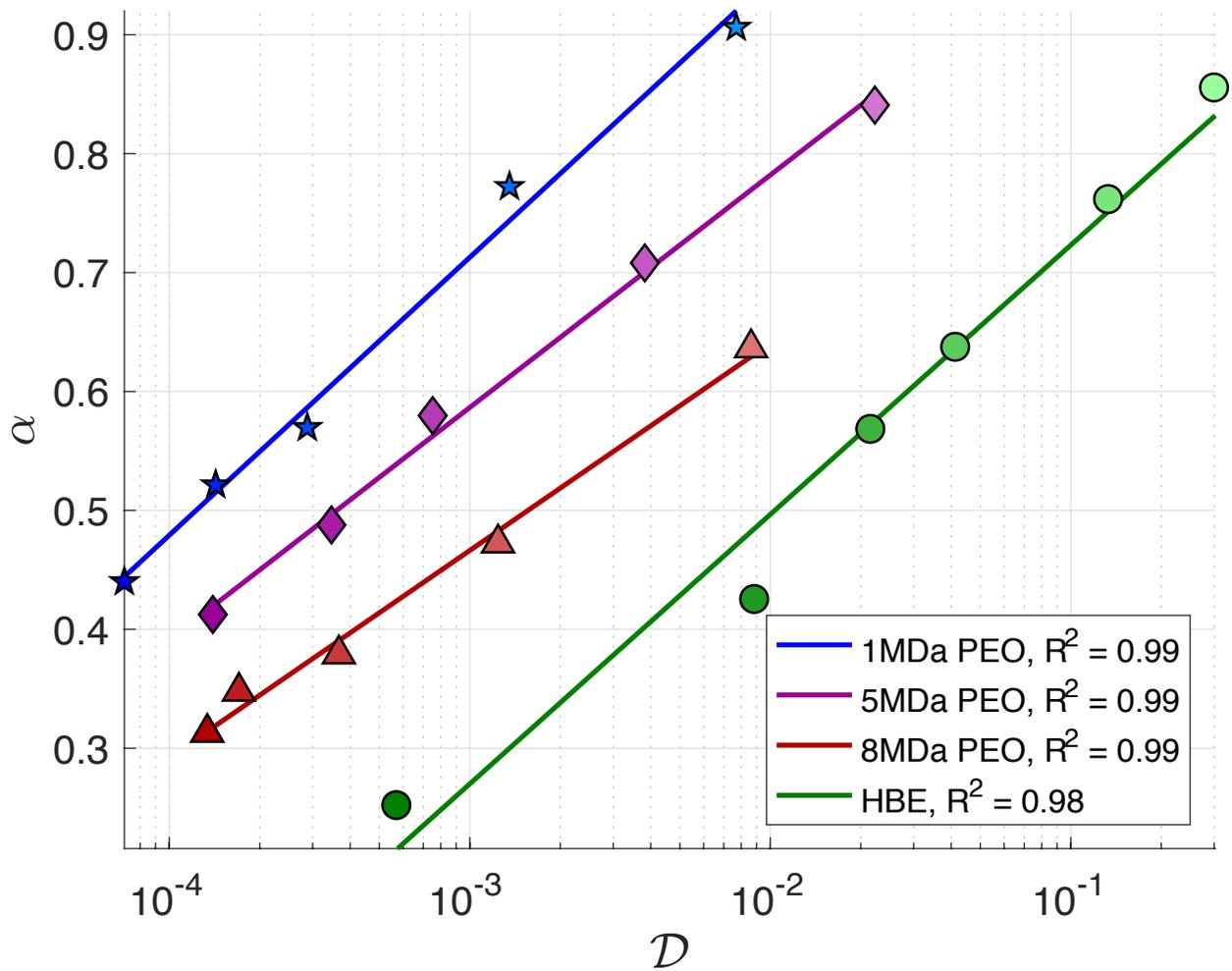

**Figure 3**: Plots of $(\log(\mathcal{D}), \alpha)$ from ensemble-averaged MSD statistics of micron diameter beads in three $M_w$ = 1, 5, and 8 MDa PEO polymeric solutions across highly entangled concentration regimes, plotted with the HBE mucus data of Figure 1.

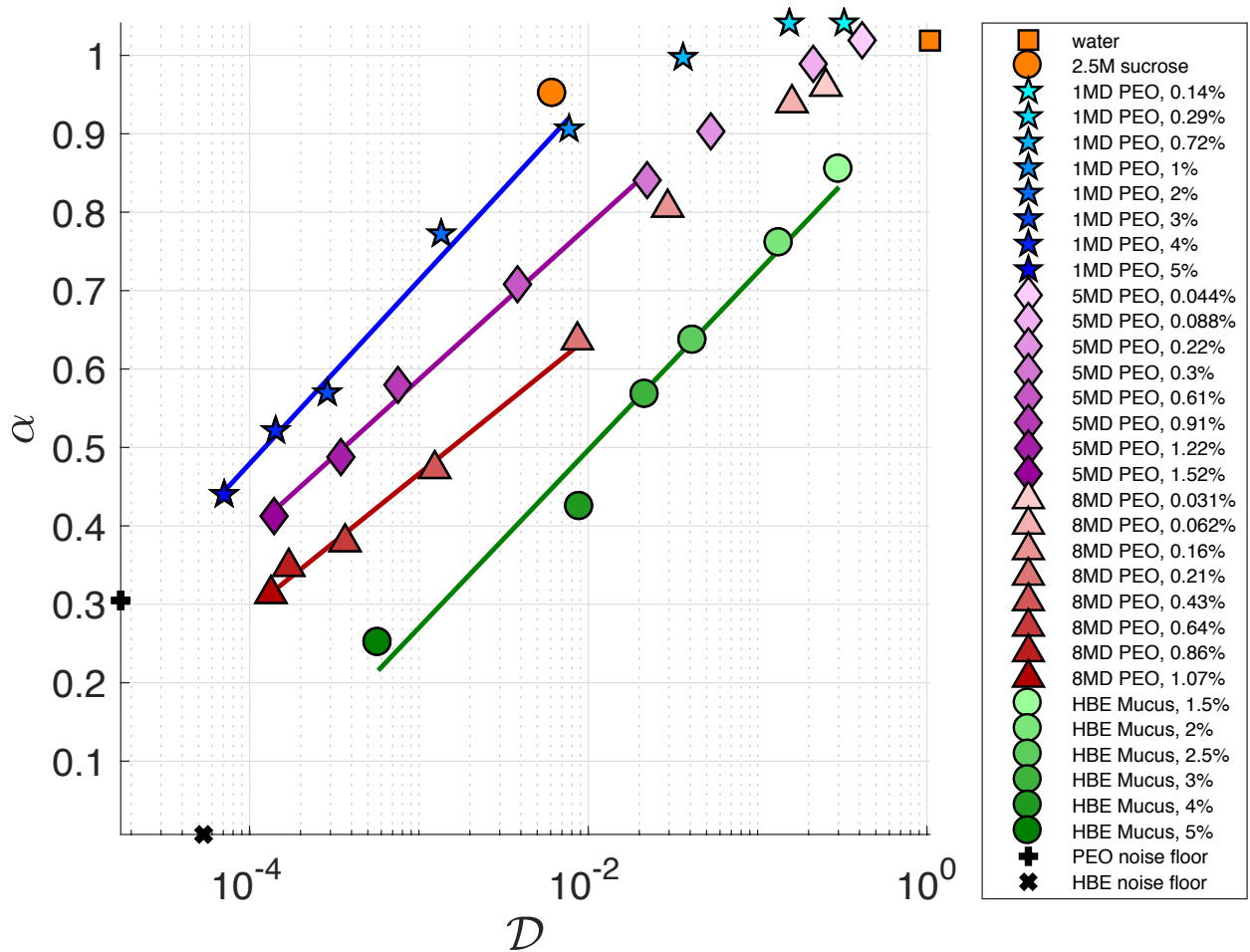

**Figure 4:** Experimental values of $(\log(\mathcal{D}), \alpha)$ for concentrations of each $M_w$ PEO and HBE mucus, both in the highly entangled regime, Figure 3, and then for each PEO as the concentration is lowered into the dilute regime where the robust scaling with concentration breaks down. As PEO concentrations approach zero, the data for each $M_w$ PEO converges to the diffusivity of water.

**Other rheological scaling relations**

Scaling relations, when validated theoretically and experimentally, are powerful tools that allow one to perform a small set of experiments to infer a small set of parameters, from which one can then extrapolate. Scaling relations for polymer solutions were a major focus of P. G. de Gennes and collaborators [24] and continue to be pursued [25] due to the power of extrapolation they provide. One illustration is the Mark-Houwink power law relation, $[\eta] = K M_w^a$, between the intrinsic viscosity, $[\eta]$, and molecular weight, $M_w$, of a polymeric solution, where $a$ and $K$ depend on properties of the *solvent and polymer*, and $a$ ranges between 0 and 1. A significant literature exists on intrinsic viscosity of polymer solutions versus solvent quality, which controls the effective volume of the molecule in solution; cf. Chapter 14 of [26] for studies of the value of $K$ for polymers in ideal or theta solvents where $a = 1/2$. In general, once $K$ and $a$ are identified for a polymer solution, the intrinsic viscosity or molecular weight can be inferred from the other, the range of intrinsic viscosities of the solution is revealed, and one can tune $M_w$ to achieve the desired values of $[\eta]$ within the available range. Like the apparent power law relations found in the present paper for HBE mucus and highly entangled PEO solutions, the Mark-Houwink relation is linear on a log-log scale, facilitating inference of the two parameters, $K$ and $a$. In our experiments and analyses, the linear fit of $(\log(\mathcal{D}), \alpha)$ versus polymer concentration *c* for PEO and wt% solids for HBE mucus involves inference of two parameters, the linear slope and intercept. Once fit, dynamic moduli are deduced across all concentrations or wt% solids in the highly entangled regime, revealing the available range of equilibrium moduli for each molecular weight PEO and wt% HBE mucus, and likewise predicting whether and at what concentration or wt% each polymer solution will undergo a viscoelastic liquid-solid transition. The power law formulas for $G'$ and $G''$ predict this viscoelastic liquid-solid transition arises at $\alpha = 1/2$, which for HBE mucus occurs between 3 and 4 wt% [23]; this prediction correlates with clinical observations of compromises in mucus clearance in lung airways as disease pathology leads to increased wt% solids in airway mucus. Note from Figures 3 and 4 that all three $M_w$ PEO solutions are predicted to undergo a transition from the viscoelastic solid regime, $\alpha < 1/2$, to the viscoelastic liquid regime, $\alpha > 1/2$.

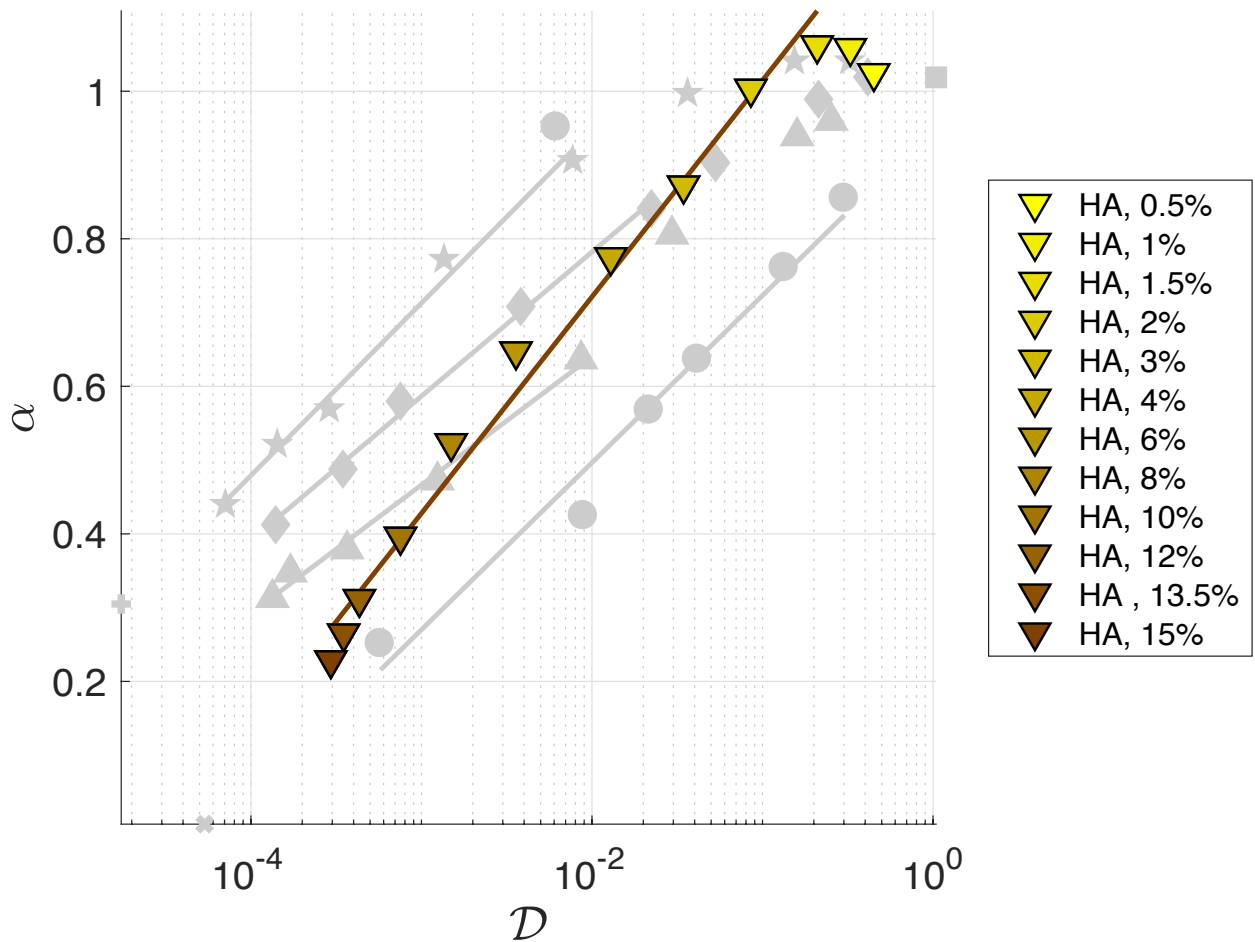

Figure 5: Overlaid on the previously shown HBE and PEO data is a range of concentrations of a single molecular weight HA. We see a distinct intercept and slope for the best-fit line through the highly entangled regime.

Several rheological signatures involving power law behavior have been proposed to characterize the viscoelastic liquid-solid transition, called the gel point for hydrogels. Dasgupta et al. proposed the point at which the temporal dependence of the relaxation modulus G(t) shifts from negative concavity to positive concavity at long time scales [27], thereby defining the gel point by $G(t) = S\, t^{-n}$, where S is the gel stiffness. Winters and Mours further defined the gel point as the transition at which tan δ is independent of frequency. Further studies have defined the gel point according to a merger of storage and loss moduli, $G' \sim G'' \sim w^n$, where n is termed the critical viscoelastic exponent [27-29]. As noted by Larsen and Furst [30], using the Mason approximation [19] which is exact for power law scaling of the MSD, as we have done in formula (3), the crossover from viscoelastic liquid to solid occurs when α = 0.5.  As noted earlier, Larsen and Furst were focused on scaling *at* the gel point, determined to be 0.6 for their solution.

**The Power Law MSD Classification Scheme:** $(\log(\mathcal{D}), \alpha)(\tau), \tau \in$ **[1/60 sec, 3 sec]**

The present study employs particle tracking of 1 micron diameter beads at 60 frames per sec for 30 seconds for all polymer solutions, thereby restricting lag times to the range [1/60 sec, 3 sec] to insure a statistically significant and independent number of observations (as discussed in [31]and in [14]).  Over this range, using Maximum Likelihood Estimation (MLE) of fractional Brownian motion in the presence of linear drift [4, 15], we find that we can reconstruct each observed bead time series with high statistical accuracy, supporting a robust, uniform, power law scaling of the stochastic process MSD over these lag times for all PEO and HA solutions, as we found previously for HBE mucus across a wide concentration range. With much higher frame rates and much longer tracking of individual beads, one would indeed observe *transient* behavior (i.e., not self-similar) in the MSD statistic, at both short timescales (ballistic inertial, viscous non-inertial, with oscillations in the transition to sub-diffusive power law) and long timescales (transition from sub-diffusive MSD power law, beyond the longest memory of the fluid sample, to viscous scaling). To gain this transient diffusive information would require both faster frame rate cameras and 3D microscopy technology (piezoelectric stage, light sheet) to track a sufficient number of particles that escape the focal plane prior to exceeding the memory timescales of the fluid.  The combination of these two timescale observations would require orders of magnitude more data storage, which for dedicated, purposeful projects is worthwhile. We note that for mucus, PEO, and HA, in the highly entangled regime, we do not observe transition to viscous scaling within the thirty second observation time. Most importantly, these sufficiently short and sufficiently long timescale MSD transient signatures, into and out of the power law regime, are not relevant for our classification scheme for highly entangled polymer solutions. Namely, we focus analysis of micron diameter beads within the lag time window of robust power law scaling. Over these lag times [1/60 sec, 3 sec], we extract the power law exponent $\alpha$ and pre-factor $D_\alpha$ (or $\mathcal{D}$) using statistical tools. First, we use the fact that fractional Brownian motion (fBm) is the unique model for self-similar, sub-diffusive, power law behavior. Second, maximum likelihood estimation methods [4, 15] in the presence of sample drift, which is typical in all the experimental data, yield efficient estimates with small errors for the fBm parameters and the deterministic drift. Validation of our fit to the experimental data is made by reconstruction of each microbead position time series from numerical simulations of the fBm + drift model using the MLE and drift parameter fits (see Figure 6 for an example). We do not need, nor do we infer, information outside of the [1/60 sec, 3 sec] lag time window, to implement this classifier scheme for power law behavior.

**Relevance to Other Microrheology Experimental Practices**

While our classification scheme focuses on what we view as a potentially important observation regarding scaling of MSD parameters, $\alpha$ and $D_\alpha$, restricted to the lag time range [1/60 sec, 3 sec] where power law behavior prevails, the experimental approach to focus on a limited range of timescales or frequencies is standard in video-based microrheology applications. Examples include the Georgiades [29] study of rheology of mucins from the GI tract, the Larsen and Furst [30] study of self-assembling peptides and polyacrylamide gels, the authors' previous studies of mucus [14, 32] and bacterial biofilms [33], the drug delivery studies reviewed by Lai and Hanes [19], and other applications reviewed by Waigh and Duncan [20, 34].

In the present study, we demonstrate that a fractional Brownian motion (fBm) model successfully captures the motion of passive probes across the time scales of the experiment. Examining data from other video microrheology studies, we find that fBm is an appropriate model for many of the previously mentioned biological studies [14, 32, 33, 35] within a limited, yet significant, band of lag times [20]. These experimental systems are not capable of reaching the high frequencies of laser tracking microrheological assays [18], thereby making the measurement of relaxation times proposed by van Zanten [36] impossible. Further, over these time scales, we do not observe α values limited to 1, 0.5, and 0 as predicted by Cai and Rubinstein [37], or 0.75 as observed in living yeast cells [38] and highly entangled f-actin solutions at intermediate time scales [39]. Our data is characterized by a wide-ranging spectrum of α's between 0.25 and 1, which is consistent with the previously cited video microrheology studies in a wide range of synthetic and biological polymer systems [14, 30, 33, 35], as well as concentrated peptide solutions [40]. Within the construct of fBm, the method of Larsen and Furst offers the most relevant method to establish the gel point of a polymeric viscoelastic solution [40] since employing GSER to these data will, by definition, always predict α=1/2 where G' and G" merge (recall formula (3)). As we previously demonstrated in HBE mucus, the GSER-based, concentration-dependent liquid-solid transition is predicted to occur near 4 wt% solids. Like the previously described studies of highly entangled peptide solutions [40], we find that the highest concentrations of large molecular weight PEO solutions exhibit a change in the concavity of the adjusted MSD curves. While Larsen and Furst warn that the critical exponent, n, at the gel point is assumed to be α = ½ within a fractional analysis, both G' and G" will scale consistently with frequency. Therefore, it is only when shifts in the adjusted MSD curves are observed that the effective liquid-solid viscoelastic transition can be observed.

Dasgupta *et al.* [25] examined microrheology of .2 MDa and .9 MDa PEO, at multiple concentrations above the overlap concentration, and with variable bead sizes. Their goal, as one of the early and seminal papers, was to illustrate, via quasi-elastic light scattering (QELS) for longer lag times and diffusing wave spectroscopy (DWS) for shorter lag times, the use of dynamic light scattering microrheometry as an alternative to macrorheometry for highly flexible, non-crosslinked, polymer solutions. The .9 MDa PEO results [25] are in the semi-dilute regime, and the 4 wt% concentration reveals a power law MSD scaling exponent of $\alpha \sim .4$, already in the viscoelastic solid regime. Our particle-tracking results on 1 MDa PEO at 3, 4, 5 wt% are most comparable to the DWS results in [25] with .9 MDa PEO at 2.2, 4, 6 wt%. Both studies show power law scaling at lag times up to 10 sec with particle tracking while at most up to .1 sec with DWS. Furthermore, the power law exponent clearly decreases proportionally with wt% in [25], indeed crossing the viscoelastic solid-liquid transition between 2.2 and 4 wt%. The authors were not focused on scaling in the MSD exponent or pre-factor with wt%, but their data reveals orders of magnitude variation in dynamic moduli with increased concentration, consistent with our scaling results, as shown for .2 MDa PEO in Figure 7 of [25]. Our particle tracking data is for 30 sec duration, and does not capture the transition to viscous scaling at sufficiently long lag times, as shown in [25] with QELS.

Larsen and Furst [40] explored *dynamics at and within the liquid-solid transition* for both physical (polyacrylamide) gels versus different crosslinking wt%, and chemical (peptide) gels at different peptide wt%.  Our studies sweep across the transition and do not attempt to investigate the power law shifts in scaling revealed by Larsen and Furst at the gel point. Nonetheless for completeness, we apply the Larsen-Furst protocol for the 8 MDa PEO solution, the only sample for which we have sufficient data above and below the inferred gel point.  Results are given in Figure 7, predicting that the gel point for this solution lies between concentrations .21 and .43, consistent with the extrapolation result from the MSD data and invoking $\alpha = 1/2$.

With regard to our studies of HBE mucus microrheology versus wt% solids [23], several other groups [35, 40, 41] have explored the rheological implications of mucus or mucin solutions versus concentration using passive microrheology. The strategy to organize the data as we have in Figure 1, showing the collinear relation of $(\log(\mathcal{D}), \alpha)$ versus wt% solids, suggests a generic scaling behavior of highly entangled polymeric solutions that has been confirmed here with three distinct molecular weight PEO solutions and HA.

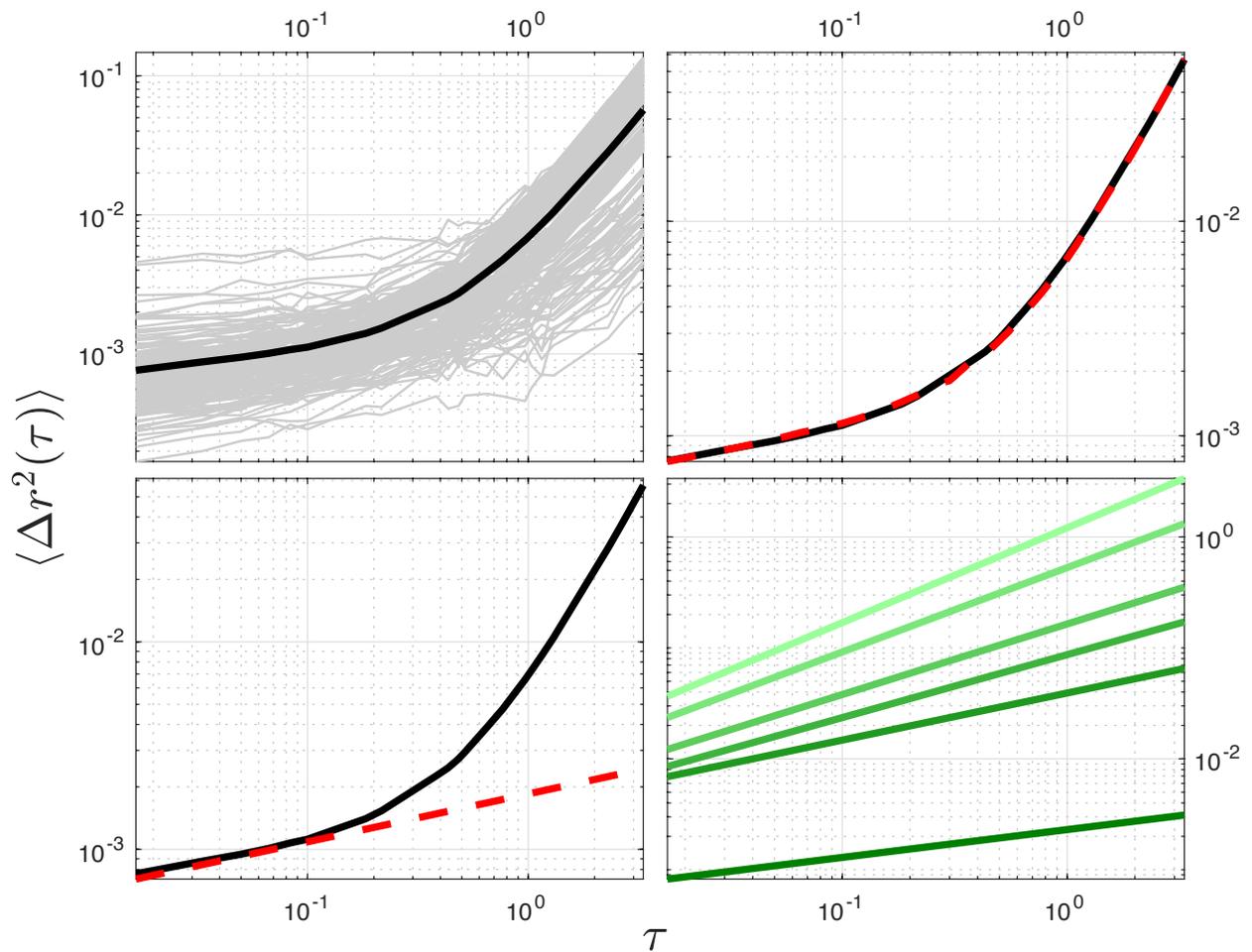

**Figure 6:** The top left panel shows individual experimental MSD curves, in gray, and the ensemble average, in black, for 5% HBE mucus on our time scale window of interest. After fitting our fBm + drift model, we reconstruct each path, and again calculate the ensemble MSD of these simulated paths. The top right panel shows the agreement between the experimental MSD (black) and the numerically reconstructed one (dotted red). Since we are interested in only the fBm parameters, we can reconstruct this purely stochastic process, shown in the bottom left panel. This procedure is repeated for each wt% HBE, and is shown as a group in the bottom right panel.

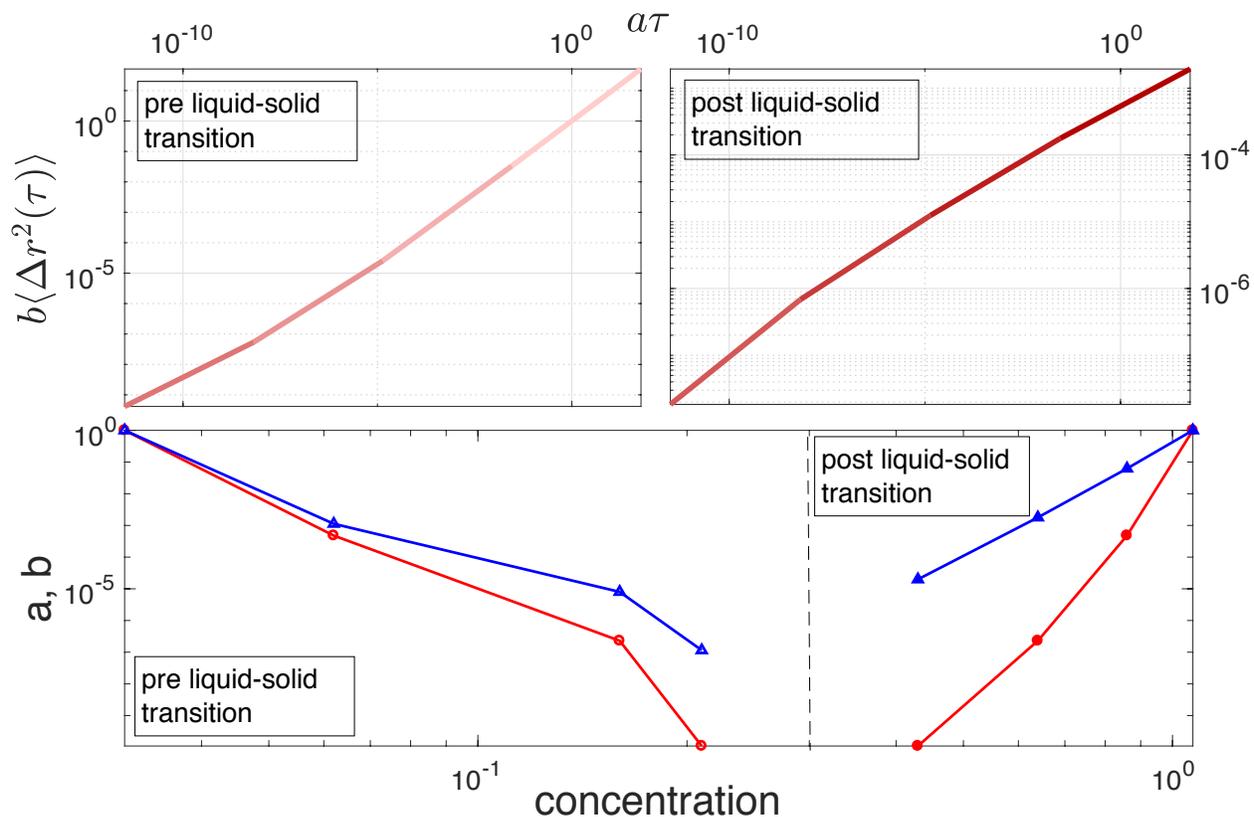

**Figure 7:** We use the Larsen-Furst protocol to investigate the viscoelastic liquid-solid transition of 8MDa PEO, above. The top two panels show so-called "master curves," created before and after the transition point, while the bottom panel shows the scaling factors, *a* and *b*, that are required to create master curves. We see trends similar to those found in polyacrylamide studied in [40].

**Conclusions**

For human bronchial epithelial mucus, synthetic highly entangled polymeric PEO at three molecular weights, and hyaluronic acid solutions, passive microbead rheology experiments and data analytics tools of our group [3, 4] point to robust scaling relations between properties of the polymeric solutions and their equilibrium microrheology. For all systems in the highly entangled regime, we find power law scaling of the MSD summary statistic of passive microbead rheology, $\langle \Delta r^2(\tau) \rangle = 2d\, D_\alpha \tau^\alpha$. The numerical pre-factor 2d, where d=2 for our particle tracking experiments, is inserted to make contact with the exact MSD scaling for fractional Brownian motion, and our inverse data analytics tools [4,17]. We non-dimensionalize $D_\alpha$ to $\mathcal{D}$ such that $\mathcal{D} = 1$ for 1 micron diameter beads in water, and show that the experimentally determined data, $(\log(\mathcal{D}), \alpha)$ versus polymer concentration, c, are collinear. Thus, concentration parameterizes the collinear data set $(\log(\mathcal{D}), \alpha)$ for each highly entangled polymeric solution.

The power law MSD summary statistic, within the lag times [1/60 s, 3 s] for each polymer system, translates via the generalized Stokes-Einstein relation to power law, self-similar, scaling of the equilibrium dynamic viscous and elastic moduli (3) within the frequency band [.33 $s^{-1}$, 60 $s^{-1}$]. This self-similar rheological behavior versus concentration, once confirmed for any given highly entangled polymeric solution, allows one to extrapolate from a discrete set of concentrations and predict the range of viscoelastic moduli across the highly entangled concentration regime. It also allows for prediction of whether a liquid-solid viscoelastic transition exists and at what concentration. A liquid-solid transition is predicted for HBE mucus at 4 wt% from Figure 1 [23], and from Figure 3, for 1, 5, and 8 MDa PEO at ~ 4%, 1.22%, and 0.43%, respectively. These results suggest a scaling relation for highly entangled polymeric solutions versus concentration, with polymer-specific properties such as molecular weight and persistence length distinguishing among the slope and intercept of the collinear data $(\log(\mathcal{D}), \alpha)$ versus concentration.

**ACKNOWLEDGEMENTS**

The authors gratefully acknowledge the support of the National Science Foundation (NSF DMS-1462992, DMS-1410047, DMS-1412844, DMS-1517274), the National Institutes of Health (NIH P300DK065988, NIH 5P41EB002025-32, NIH/NHLBI R01-HL116228), the Natural Sciences and Engineering Research Council of Canada (NSERC RGPIN-2014-04225), and the Cystic Fibrosis Foundation (CFF BOUCHER15R0, CFF HILL16XX0, and CFF RAMSEY16I0).

**METHODS**

**Data Analytics**

Each tracked microbead results in a position time series of $N$ observations (1800 for all data in this study) in each coordinate. Observation of particle trajectories and the power-law behavior of empirical mean-squared displacement curves over the observed timescales support modeling each trajectory as fractional Brownian motion (fBm), while allowing for constant linear drift in each coordinate [15]. The presence of persistent drift in a tracked microbead introduces drift-dependent uncertainty and error (in the MSD of each trajectory, in the ensemble-averaged MSD, and in the inferred $G^*$) if one does not account for drift. If one does estimate drift and then subtract it from the particle increments, significant distortions arise at log lag times. In [15] we describe the advantages of maximum likelihood estimation of the two fBm parameters, $\alpha$ and $D_\alpha$, and the drift velocity, $\mu$, in each coordinate, performed *simultaneously* and not sequentially. Here, we simply apply the methods in [15], noting that this is an accurate and robust method to extract the sub-diffusive parameters $\alpha$ and $D_\alpha$.

Since we assume the motions in the *x* and *y* directions are independent, the following formulation describes the 1-dimensional protocol for modeling the position time series for each coordinate. If the position at time *t* of a single path is denoted $X(t)$, then the position process is assumed to be of the form

$$X(t) = \mu t + \sqrt{2D}W_\alpha(t) \qquad (4)$$

where $W_\alpha(t)$ is a continuous Gaussian process with mean zero and covariance

$$\text{cov}(W_\alpha(t), W_\alpha(s)) = \frac{1}{2}(|t|^\alpha + |s|^\alpha - |t-s|^\alpha), \qquad 0 < \alpha < 2 \qquad (5)$$

The stochastic process uniquely describes fractional Brownian motion with two parameters, $\alpha$ and $D_\alpha$. Given this model, we use maximum likelihood estimation as detailed in [15] to simultaneously estimate the three model parameters, $\mu$, $\alpha$, and $D_\alpha$, directly from the position time series data. This method accounts for deterministic drift and returns the parameters of the stochastic process, $\alpha$ and $D_\alpha$, *for each particle trajectory*.

Next, we want to ensemble average the stochastic parameters for each microbead trajectory, having successfully reduced errors due to typical variation in persistent motion across the ensemble of tracked beads. (Note: one cannot ensemble average the bead increments versus lag time, since they are distorted by drift, increasingly so with increased lag times.) The ensemble averaging of fBm parameters, however, *requires* non-dimensionalization of the prefactor $D_\alpha$, which has physical units that scale with $\alpha$. That is, one cannot average individual values of $D_\alpha$ unless they have the same $\alpha$ value, which essentially never happens unless the fluid is known to be purely viscous. To remedy this, we propose the following method for non-dimensionalization of $D_\alpha$, giving a new parameter, $\mathcal{D}$, that can be used to perform ensemble averaging.

The pre-factor $D_\alpha$ has units $\mu m^2 / s^\alpha$. In order to non-dimensionalize, we normalize $D_\alpha$ by a chosen factor with the same units. Since we are focusing on polymeric solutions where the pure solvent is water, we choose the pre-factor such that the non-dimensional pre-factor, $\mathcal{D}$, will be 1 for 1 micron diameter beads in water. Recall the Stokes-Einstein relation for the diffusivity of water, $D_w = k_B T / 3\pi d \eta_w$, where $\eta_w$ is the viscosity of water, $k_B$ is the Boltzmann constant, $T$ is temperature, and $d$ is the bead diameter. If we divide by $D_w^\alpha$, that accounts for the $s^\alpha$ units, and then we use the bead diameter d raised to the appropriate power to arrive at a dimensionless pre-factor, $\mathcal{D}$:

$$\mathcal{D} = \frac{d^{2(\alpha-1)} D_\alpha}{D_w^\alpha}. \tag{6}$$

Clearly, $\mathcal{D} = 1$ for 1 micron diameter beads in water.

For each sample, we now have a collection of points in $(\mathcal{D}, \alpha)$ space corresponding to the number of particles tracked. Observation consistently reveals that $\alpha$ is fit well by a normal distribution, and that $\mathcal{D}$ is well fit by a lognormal distribution.

*Thus, we report expected values and variances for each parameter corresponding to their respective distributions*. This evidence explains why the $(\mathcal{D}, \alpha)$ data for all samples are presented as semi-log plots, $(\log(\mathcal{D}), \alpha)$.

**Statistic of Linear Regressions**

For each molecular weight PEO and HBE mucus, a simple linear regression of the form $\alpha = \mu + \beta \log(\mathcal{D})$ was fit to the $(\log(\mathcal{D}), \alpha)$ pairs that correspond to different concentrations, *c*. The $R^2$ value for each dataset can be interpreted as the percentage of variance in $\alpha$ that is explained by the fit trend line. Next, 95% confidence intervals around the fit parameters, $\mu$ and $\beta$, are calculated for each trend line by using estimates of the standard errors for each parameter and corresponding percentiles of the *t*-distribution. Comparisons of the slopes, $\beta$, Figure 8, and the intercepts, $\mu$, Figure 9, across each molecular weight PEO and HBE mucus can now be made.

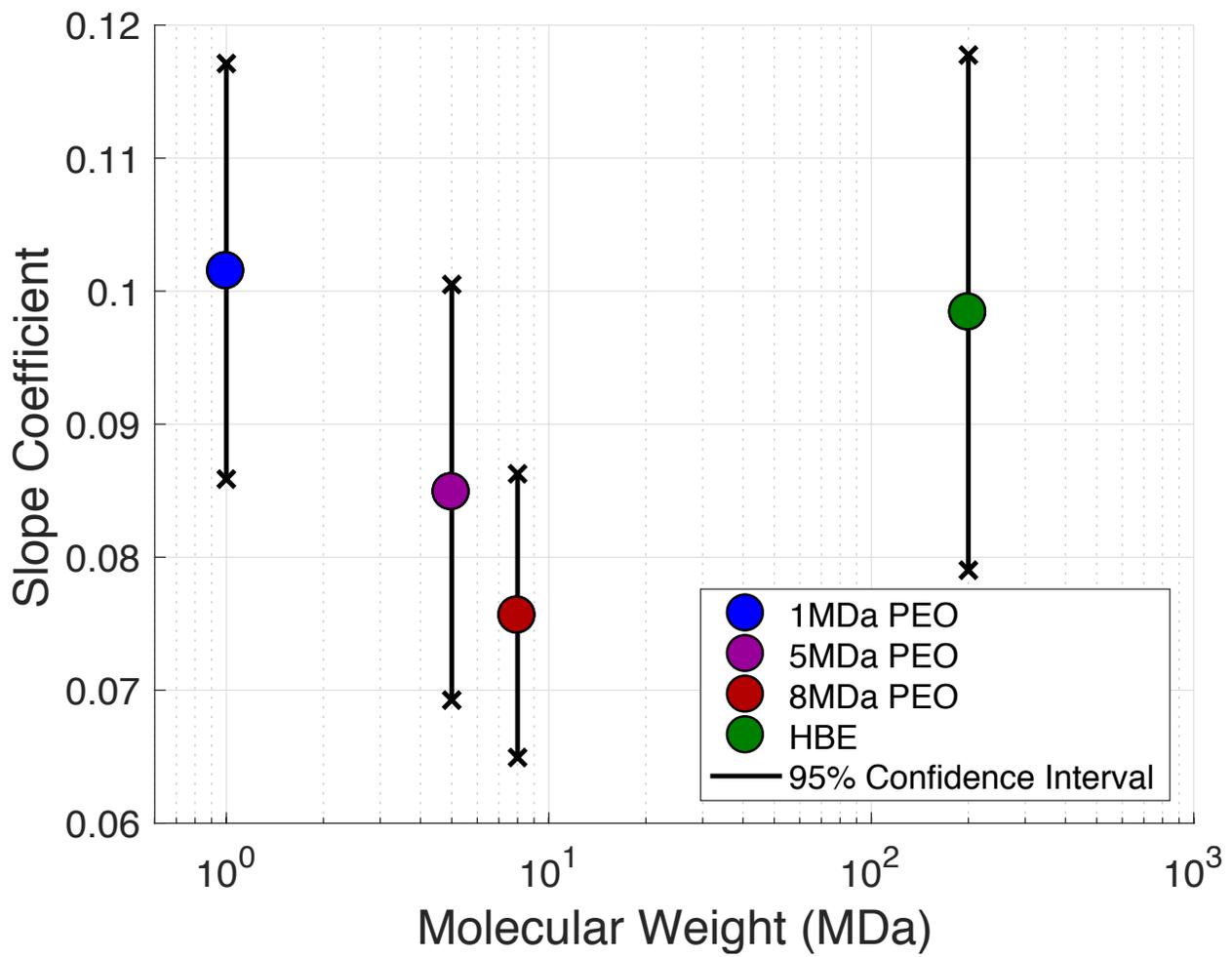

**Figure 8**: Each trend line follows the form $\alpha = \mu + \beta \log(\mathcal{D})$. Shown above are the estimated values and 95% confidence intervals for the $\beta$'s, or the slope coefficients, from each trend line. A weak trend is apparent as molecular weight increases.

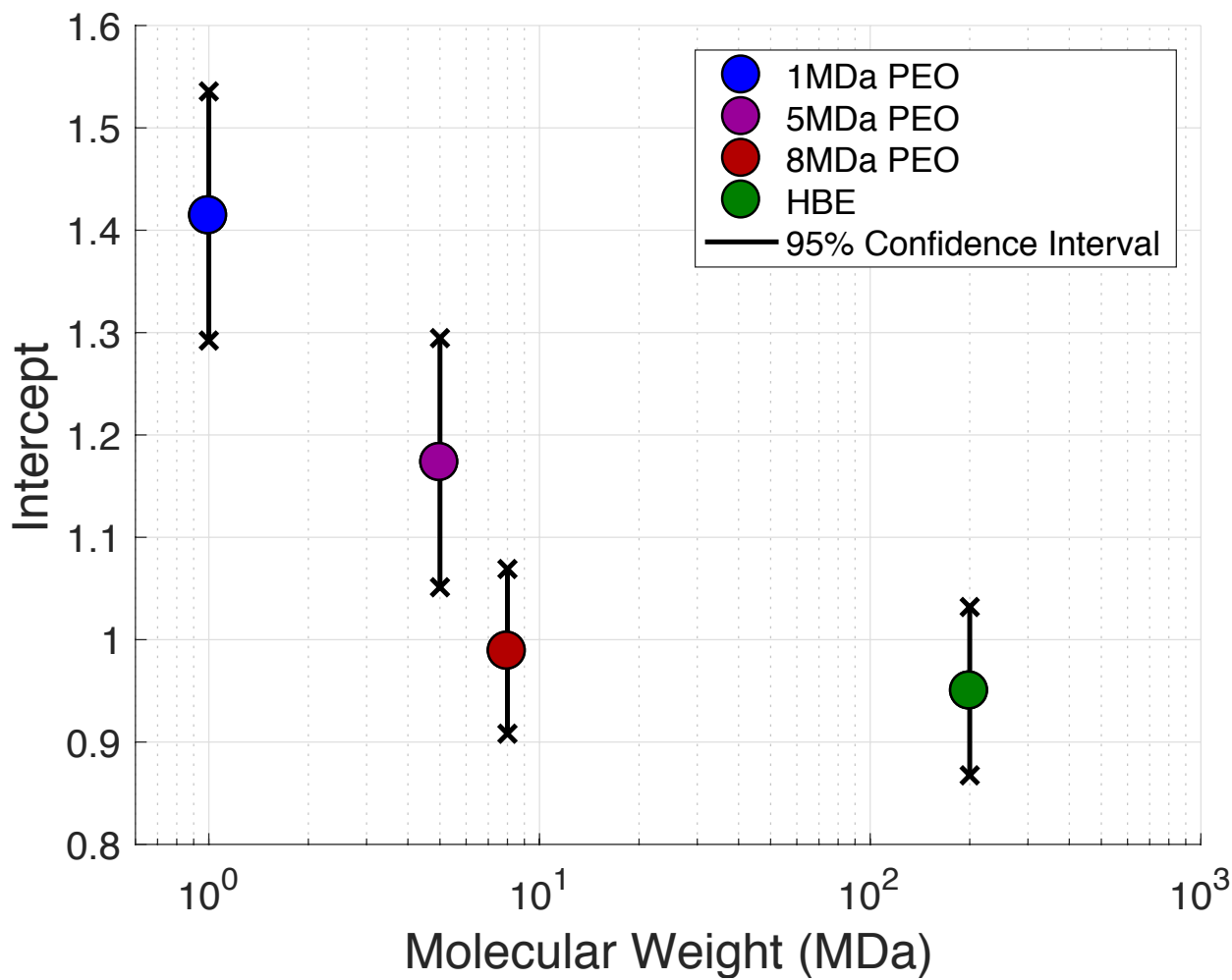

**Figure 9:** The estimated values and 95% confidence intervals for the $\mu'$s, or intercepts, in the best-fit lines for PEO and HBE mucus. Here, we observe statistically different values and evidence for a logarithmic scaling of the intercept with molecular weight M, in units of MDa, for PEO, with the extremely high M, multi-species, polydisperse HBE mucus inserted only for comparative purposes.

**Polymer Physics of Synthetic, Polyethylene Oxide (PEO), Hydrogels**

We select polyethylene oxide (PEO) to synthesize polymeric solutions, since PEO chains are linear and non-cross-linked, available in a wide range of molecular weights, and biologically inert. PEO has a persistence length of ~0.93 nm; we refer to [42] for detailed studies of mesh size of the polymer concentration, polymer radius of gyration, polymer molecular weight, and overlap concentration. We select a range of molecular weights, 1, 5, and 8 MDa, as specified by the supplier, and estimate the highly entangled concentration per $M_w$ to be 1.45, 0.44, and 0.31 mg/mL, respectively. The data in Figure 3 is above these concentrations, while Figure 4 shows data in the dilute concentration limit for each $M_w$ PEO.

**Solution Preparation**

Each solution was prepared by weighing out approximately 20g of deionized water containing 1:500 dilution of 2% 1$\mu$m diameter yellow-green (YG) microspheres (Fluorospheres, Molecular probes) in a 50mL conical centrifuge tube. A quantity of PEO, which corresponds both to the target molecular weight and overlap parameter, is also weighed out. The PEO is added to the DI water and beads while stirring vigorously on a standard lab vortexer. Once the solute is well wetted, the centrifuge tube is capped and sealed with Parafilm and left to mix on a slow-rotator for a minimum of 48 hours before performing microbead tracking.

**Overlap Parameter**

Since the overlap concentration is a means to normalize polymer solutions based on polymer-polymer interactions, the overlap parameter $P$ is defined as the ratio of the concentration of a test solution to that solution's expected overlap concentration, $P = c/c^*$. As long as the units and solution specification for $c$ and $c^*$ are identical, this normalization relationship for $P$ should hold.

**Using Mass Fraction for Concentration**

The mass of the solute becomes a more significant proportion of the total solution's mass and volume at $P \times c^*$ concentrations with higher $P$, especially for polymers that have lower molecular weight. This discrepancy can introduce a systematic error when defining traditional weight-per-volume concentrations during sample preparation. To minimize this error, we measured the mass of both the solute and solvent used to make each solution and computed the mass fraction as $M_{solute}/(M_{solute} + M_{solvent})$. Converting $c^*$ from units of mg/mL to unitless mass fraction results in virtually identical numerical quantities (small error), since the contribution of solute to total solution mass is quite small at overlap, and the density of water is 0.997 g/cm$^3$ at room temperature.

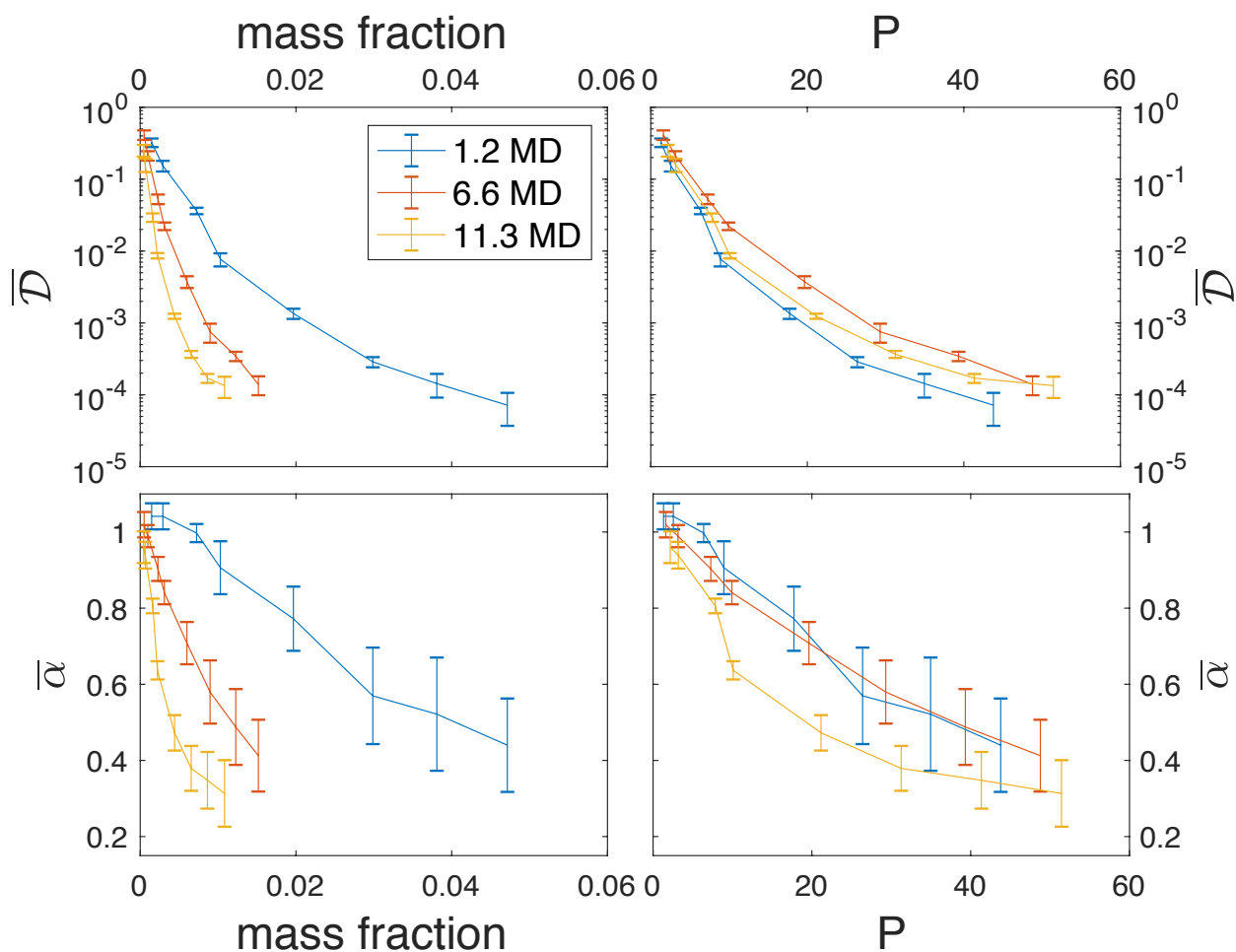

**Figure 10**: Shown above are mass fractions and overlap parameters plotted against mean values of $\mathcal{D}$ and $\alpha$ per concentration for each molecular weight PEO. These molecular weights were determined with light scattering measurements and vary somewhat from reported values on the bottle.

**Sample Preparation and Data Collection**

A silicone spacer (Grace Bio-labs 654008), mounted onto a 24x40x1 ½ coverslip, served as the sample chamber for PEO solutions. After each solution rotated for at least 48 hours, $2\mu L$ was delivered to the sample chamber through the use of a positive-displacement pipette (Gilson Microman M25). A second coverslip sealed the sample chamber. Equilibrating 15-30 minutes before collecting video data reduces but does not totally eliminate sample drift, which our maximum likelihood estimation disentangles from the stochastic fluctuations.

Camera exposure times were tested for oversaturating the CCD and frames were collected once an exposure time of 10 ms was found to satisfy the desired condition of high SNR with no oversaturation, since oversaturating pixels reduces tracking fidelity in Video Spot Tracker. Combining a read-out delay of 18.7 ms per camera frame results in a 40 frames per second collection rate. The duration of the collected videos range from 1 minute for water to 5 minutes for the highest PEO concentrations. The scaling factor to convert pixel sizes to physical dimensions for the 40X objective was 0.157 $\mu$m/pixel. The fluorescent microbeads were tracked by the Video Spot Tracker software as previously described [23, 43] (http://cismm.web.unc.edu/).

**Molecular weight determination**

The ($M_w$) and radius of gyration ($R_g$) of mucus, PEO, and HA were assessed by differential refractometry [44]. A 500$\mu$l sample is chromatographed on a Sepharose S1000 column (Amersham Pharmacia), which was eluted with 200mM sodium chloride with 10mM EDTA at a flow rate of 0.5 ml/min. The column effluent was passed through an in–line Dawn EOS laser photometer coupled to a Wyatt/Optilab DSP inferometric refractometer to measure light scattering and sample concentration, respectively. Mucus $M_w$ and $R_g$ are determined by best fit of scattering data at multiple angles to a Berry model [45]. We determine the measured molecular weight of PEO molecules to be larger than reported by the manufacturer (Figure 10).